\documentclass[twocolumn,amssymb,fleqn]{revtex4-2} 
\usepackage{epsfig,amssymb,amsmath,graphicx,hyperref,subfigure}  
\usepackage{color}

\usepackage{multirow} 
\usepackage{tabularx}

\usepackage{graphicx}
\graphicspath{{img/}}

\newcommand{\be} {\begin{eqnarray}}
\newcommand{\ee} {\end{eqnarray} }

\newcommand{\f} {\frac }
\newcommand{\la} {\langle }
\newcommand{\ra} {\rangle }

\begin{document}

\title{Emergent dynamic stress regulators via coordinated thermal fluctuations \\ and
stress in harmonic crystalline lattices}
\author{Zhenwei Yao}
\email{zyao@sjtu.edu.cn}
\affiliation{School of Physics and Astronomy, and Institute of Natural
Sciences, Shanghai Jiao Tong University, Shanghai 200240, China}
\begin{abstract} 
Understanding thermal fluctuations yields insights into a wide range of
  behaviors in many-body systems. In this work, we analyze the dynamical
  adaptation of two-dimensional crystalline lattice system under harmonic
  interaction in response to the intricate interplay of thermal agitation and
  mechanical stress by developing the characteristic stress-absorbing quadrupole
  structures and stress-releasing fold structures. These thermally driven stress
  regulator structures serve as a tangible embodiment of thermal fluctuations,
  offering a unique perspective on the characterization and manipulation of the
  elusive fluctuations.  Specifically, we reveal the stretch-driven alignment
  and linear accumulation of quadrupoles, characterize the formation and
  proliferation of folds, and present the phase diagram of the dynamical states
  defined by these characteristic structures.  This work demonstrates the
  promising avenue of re-examining classical mechanical systems subject to
  thermal agitation, which is of fundamental physical interest and has potential
  practical significance in the design of mechanical devices in thermal
  environments. 
\end{abstract}

\maketitle

\section{Introduction}

Thermal fluctuation is a fundamental phenomenon strongly connected to a wide
range of behaviors in many-body systems at finite temperature, ranging from
the stability of the equilibrium
state~\cite{callen1951irreversibility,prigogine1971,ehrenfest2002conceptual} to
the emergence of exceedingly rich spatiotemporal structures spanning multiple
orders of magnitude in both time and length scales in extensive physical and
biological
systems~\cite{Nelson2004c,galliano2023two,toner1995long,vasseur2014increased}.
Of particular interest is the intermediate regime, where thermal excitations
have not completely dominated the interactions of the constituents of the
system. In this regime, the thermal fluctuation is capable of exciting inherent
characteristic structures that are inaccessible or disintegrated at either low
or high temperatures. As manifestations of thermal fluctuations, these excited
structures offer a unique perspective on the characterization and manipulation
of fluctuations, complementing the traditional analyses of fluctuations from the
aspects of
thermodynamics~\cite{callen1951,kubo1966,prigogine1971,mcclure2021thermodynamics},
correlation function~\cite{zwanzig1965time,Landau1999a}, and renormalization
group~\cite{david1988crumpling,NelsonP2016}.

The two-dimensional crystalline lattice system serves as an ideal model for
exploring the fundamental structures excited by the interplay of thermal
fluctuations
and mechanical stress. Mechanical inquiry into the stressed crystalline sheet
system at zero temperature has revealed a series of elastic and plastic
deformation
patterns~\cite{audoly2010elasticity,vernizzi2011platonic,bowick2016colloidal,soft_book}
ranging from wrinkles and
folds~\cite{cerda2002wrinkling,holmes2010draping,tallinen2008deterministic,Grason2013}
to topological defects and
fractures~\cite{bowick2009two,wales2014chemistry,Mitchell2016,yao2019command,sun2025defect},
all of which enable the system to adapt to curved spaces or external mechanical
constraints. The behavior of crystalline sheet is even richer in thermal
bath~\cite{keim2004harmonic,paulose2012fluctuating,horn2014does}.  For example,
thermal fluctuation leads to remarkable enhanced bending
rigidity~\cite{Nelson2004c,blees2015graphene,NelsonP2016,wan2017thermal,bowick2017non}
and featured thermal buckling
transitions~\cite{le2021thermal,shankar2021thermalized,morshedifard2021buckling,hanakata2021thermal,chen2022spontaneous},
which profoundly enrich the subject of classical
elasticity~\cite{Landau1986,audoly2010elasticity} and hold the promising
possibility for applications in mechanical nanodevices designed to operate in
thermal environments~\cite{ekinci2005nanoelectromechanical}.

In this work, we explore the microscopic dynamics of the stressed
two-dimensional crystalline lattice system under harmonic interaction subject to
thermal fluctuation.  Specifically, our model consists of a triangular lattice
of identical linear springs that wraps seamlessly around a cylindrical
substrate to enforce periodic boundary condition, as illustrated in
Fig.~\ref{model}(a).  We create a thermal environment by perturbing the particle
configuration away from mechanical equilibrium; the subsequent evolution of the
system conforms to Hamiltonian dynamics.  The numerical experimental approach
enables us to capture the intricate dynamics and reveal the characteristic
quadrupole and fold structures that emerge from the coordinated effects of
thermal agitation and mechanical stress.

A quadrupole consists of four disclinations of opposite signs (analogous to
electric charges) organized in a square configuration~\cite{nelson2002defects}.
As a thermal excitation, the quadrupole structure originates from local
fluctuations in particle positions. We observe the stretch-driven alignment
and linear accumulation of quadrupoles forming transient shear band structures.
Fold structures, on the other hand, represent a prevalent deformation mode in
both compressed and stretched lattices under strong thermal agitation; the
proliferation of folds ultimately triggers the collapse transition of the
lattice. These excited structures as stress regulators define the
distinct dynamical states at varying levels of temperature and stress. This work
suggests the avenue of re-examining mechanical systems in thermal environment,
and especially of exploring the thermal physics of stress regulators,
whose roles in mechanical systems have long been
studied~\cite{Landau1986,audoly2010elasticity}.

\section{Model and Method}

\begin{figure}[t]  
\centering 
\includegraphics[width=3.4in]{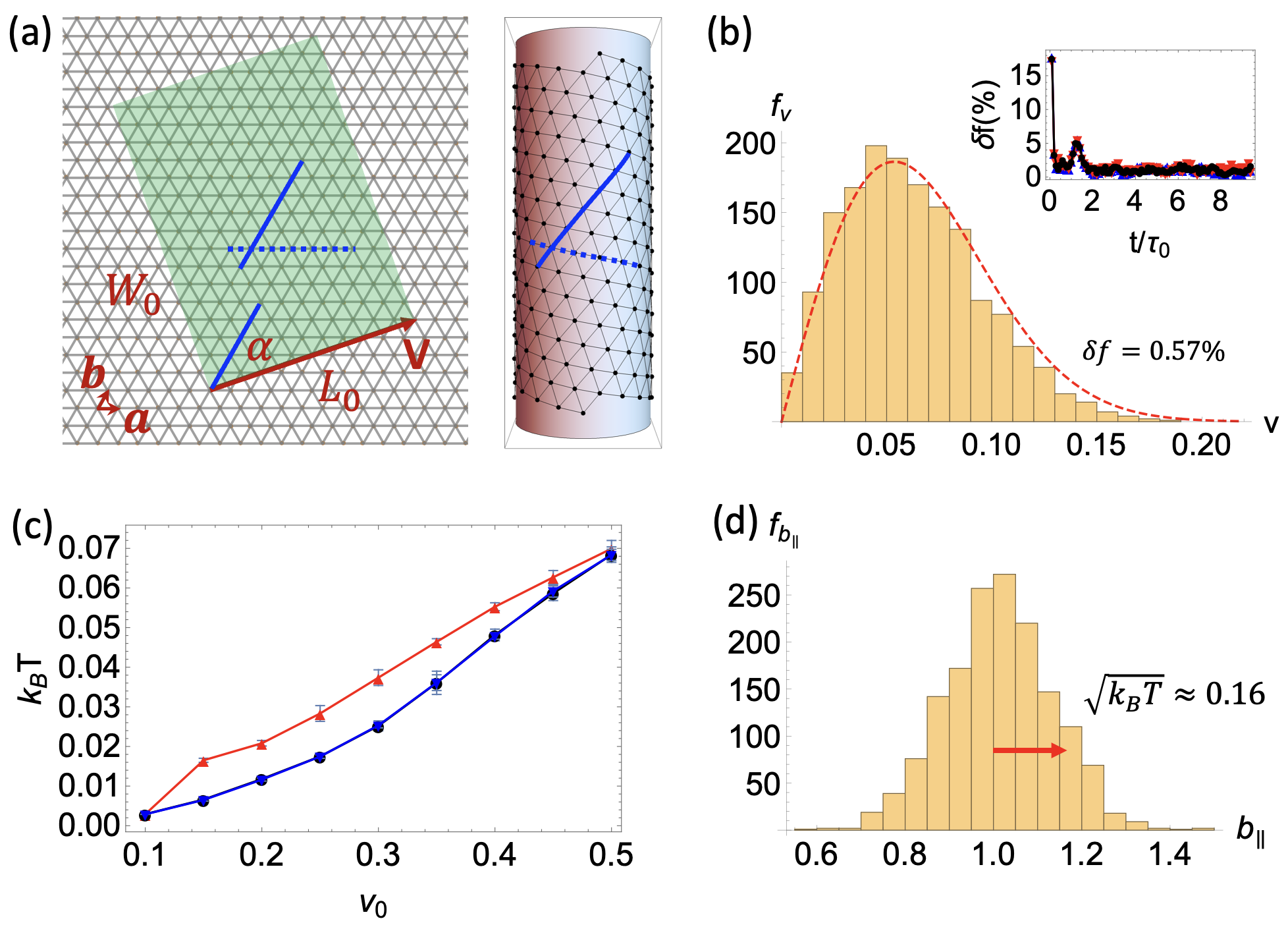}
  \caption{Thermalized crystalline lattice confined on the cylindrical surface.
  (a) Schematic plot of the lattice-on-cylinder model. The lattice is generated
  by the periodicity vector $\vec{V}=p \vec{a} + q \vec{b}$. $(p,q)=(8, 4)$
  (left) and $(16, 8)$ (right). The $x$ axis is along the $\vec{V}$ vector.
  (b)-(d) show the thermalization of the disturbed lattice system. (b) The
  convergence of the speed distribution towards the Maxwell distribution (dashed
  red curve) at $v_0=0.1$. The inset figure shows the rapid convergence process
  as characterized by $\delta f$ (the deviation from the Maxwell distribution)
  at $v_0=0.1$ (blue), $0.25$ (red) and $0.5$ (black). (c) Plot of temperature
  (as derived from the Maxwell distribution) versus the initial disturbance
  strength $v_0$ at $\epsilon_0=-20\%$ (red), $0$ (blue) and $20\%$ (black).
  (d) Instantaneous distribution of the length of bonds parallel to the $x$
  axis. The magnitude of the dispersion derived in theory is indicated by the
  red arrow. $v_0=0.3$. $t=25,000\tau_0$ and $\epsilon_0=0$ [(b) and (d)].
  }
\label{model}
\end{figure}

In our model, the rectangular crystalline lattice of length $L_0$ and width
$W_0$ is generated by the periodicity vector $\vec{V}$ connecting a pair of
lattice points; see Fig.~\ref{model}(a)~\cite{mughal2012dense,mughal2014theory}.
The lattice spacing is $\ell$.  $\vec{V}=p \vec{a} + q \vec{b}$, where $\vec{a}$
and $\vec{b}$ are the elementary lattice vectors. $p$ and $q$ are
positive integers.  $L_0=|\vec{V}|$.  The rectangular lattice can seamlessly
wrap a cylinder of radius $R_0$; $L_0=2\pi R_0$. The $x$ axis is along the
$\vec{V}$ vector. The tilt angle of the lattice with respect to the $x$ axis is
denoted as $\alpha$; $\alpha \in [0^{\circ}, 60^{\circ})$. The particles in
the lattice are bonded by identical linear springs of stiffness $k_0$ and rest
length $\ell_0$. The cylindrical substrate serves as a geometric constraint
only; no friction is involved. Due to the nearest-neighbor interaction and the
local flatness of the cylindrical manifold, the curvature effect of the
cylindrical substrate can be neglected. Therefore, we are essentially
dealing with a flat crystalline lattice that is periodic in one direction. The
edges of the lattice are stress free. The initial horizontal strain is
specified by the lattice spacing $\ell$: $\epsilon_0 =
(\ell\cos\alpha-\ell_0\cos\alpha)/(\ell_0\cos\alpha)=(\ell-\ell_0)/\ell_0$,
where $\alpha$ is the tilt angle of the crystalline lattice.

The crystalline lattice on the cylinder is then mechanically relaxed to the
lowest energy state by the steepest descent method. Specifically, we first
calculate the total force on each particle arising from the particle-particle
interaction. The tangential force on particle $i$ is denoted as $\vec{F}^{(i)}$, where
$i=1,2,3...N$. $\vec{F}^{(i)}= F^{(i)}_{\varphi} \hat{e}_{\varphi}  +
F^{(i)}_{z} \hat{e}_{z}$, where $F^{(i)}_{\varphi}$ and $F^{(i)}_{z}$ are the
$\varphi$- and $z$-components of the force.  The positions of the
particles are characterized by the cylindrical coordinates $(\varphi^{(i)},
z^{(i)})$, and they are collectively updated as follows: 
\be
\varphi^{(i)} &=& \varphi^{(i)}+\tilde{F}^{(i)}_{\varphi} s, \nonumber \\
z^{(i)} &=& z^{(i)}+\tilde{F}^{(i)}_{z} s,
\ee
where $\tilde{F}^{(i)}_{\varphi}=F^{(i)}_{\varphi}/F_{\varphi,max}$, and
$\tilde{F}^{(i)}_{z}=F^{(i)}_{z}/F_{z,max}$. $F_{\varphi,max}$ and
$F_{z,max}$ are the maximum tangential forces on the particles. The step
size $s$ is reduced from $s=10^{-3}$ to $s=10^{-5}$ in the entire relaxation
process to ensure that the maximum residual force on the particles is at the
order of $10^{-6}$ or even lower after a few million simulation steps. In this
force-driving protocol, each movement of the particles contributes to the
reduction of the energy. The mechanical relaxation leads to the shrinking or
expansion of the crystalline lattice along the $z$ axis, depending on the
value of the initial lattice spacing $\ell$. Even when $\ell=\ell_0$, the
crystalline lattice contracts slightly as a consequence of the line-tension
effect.

The disturbance is imposed on the mechanically relaxed lattice by specifying an
initial velocity $\vec{v}^{ini}_i$ on each particle $i$~\cite{supp_thermal}. In
cylindrical coordinates $(\rho, \varphi, z)$, $\vec{v}^{ini}_i= v_0 (\cos \phi
\hat{e}_{\varphi} +  \sin \phi \hat{e}_{z})$, where $\phi$ is a uniform random
variable in the range of $\phi \in [0, 2\pi)$.  The constant speed $v_0$
indicates the strength of the initial disturbance. The subsequent evolution of
the system conforms to Hamiltonian dynamics. The Hamiltonian of the system is 
\be
H = \sum_{i=1}^{N} \f{p_i^2}{2m_0} + \sum_{\la ij \ra}
\f{1}{2}k_0(r_{ij}-\ell_0)^2, \label{m1}
\ee
where $p_i$ and $m_0$ are the momentum and mass of particle $i$, and $r_{ij}$ is
the length of the bond $\la ij \ra$. In our model, the lattice permits
self-intersection, meaning that bonds and particles can pass through each other
without incurring any energetic penalty.

The equations of motion are 
\be
m_0 (\ddot{\rho} - \rho \dot{\varphi}^2)  &=& F_{\rho}^{int}+F_{\rho}^{ext},
\label{a1}   \\
m_0 (2\dot{\rho} \dot{\varphi} +  \rho \ddot{\varphi}) &=& F_{\varphi}^{int},
\label{a2}\\
m_0 \ddot{z} &=& F_z^{int}, \label{a3}
\ee
where $F_{\rho}$, $F_{\varphi}$ and $F_{z}$ are the $\rho$-, $\varphi$- and
$z$-components of the force on the particle concerned. The superscripts $int$ and
$ext$ indicate the internal interaction force and the external force.  The
external force imposed by the frictionless cylindrical substrate is normal to the
surface; no external force is thus involved in Eqs.(\ref{a2}) and (\ref{a3}).  Since
the motion of the particles is confined on the cylindrical surface, $\rho$ is a
constant; $\rho=R_0$. Both the $\ddot{\rho}$ term in Eq.(\ref{a1}) and the
$2\dot{\rho} \dot{\varphi}$ term in Eq.(\ref{a2}) automatically vanish.
Eq.(\ref{a1}) is also automatically satisfied under the combined centripetal
force $F_{\rho}^{int}+ F_{\rho}^{ext}$.  Consequently, the three equations of
motion are reduced to the two equations of Eqs.(\ref{a2}) and (\ref{a3}), with
the term $2\dot{\rho} \dot{\varphi}$ in Eq.(\ref{a2}) being zero.

The equations of motion are numerically integrated by the Verlet
method~\cite{rapaport2004art,supp_thermal}.  We denote the position of the
particle $(\varphi, z)$ by $\vec{x}$. According to the Verlet algorithm, the
update of the particle position conforms to the following recurrence relation:
\be
\vec{x}_i(t+2h) = 2 \vec{x}_i(t+h) - \vec{x}_i(t) +
\ddot{\vec{x}}_i(t+h)h^2   +
{\cal{O}} (h^4).\nonumber
\ee
where the position at $t+2h$ is obtained from the positions at $t$ and $t+h$. $h$ is
the time step; $h=10^{-3}$ in simulations.  In the cylindrical system, the
external force that is perpendicular to the surface does zero work to the
collection of the particles whose motion is confined on the surface. Therefore,
the total mechanical energy of the system is conserved during its dynamical
evolution. In simulations, the energy is well conserved. After ten million
simulation steps at the time step of $h=10^{-3}$, the relative variation of the
total mechanical energy is at the order of $0.01\%$ within the explored range of
relevant parameters.

In this work, the units of length, time and mass are $\ell_0$,
$\tau_0$ and $m_0$, where $\tau_0=\sqrt{m_0/k_0}$.  The key control parameters
are $\epsilon_0$ (initial horizontal strain) and $v_0$ (initial speed).

\begin{figure*}[t]  
\centering 
\includegraphics[width=7in]{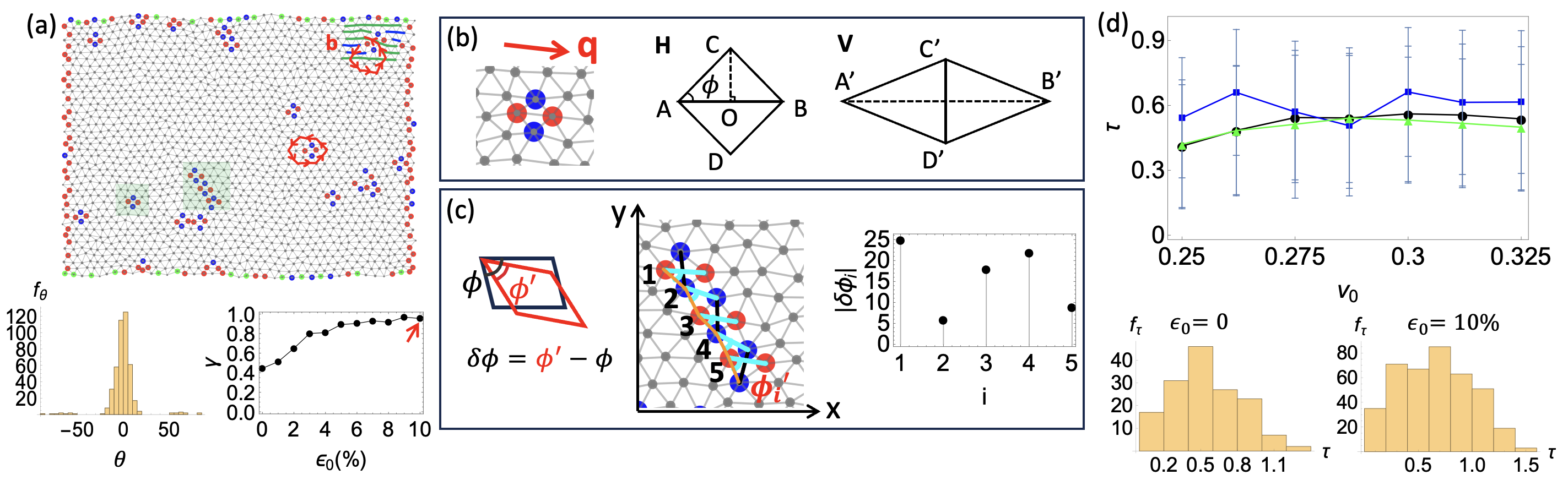}
  \caption{Characterization of thermally driven quadrupoles in a typical stretched
  crystalline lattice. (a) Emergence of quadrupoles by the combined effects of
  stress and thermal fluctuation. A quadrupole consists of four disclinations of
  opposite signs organized in a square configuration; the five- and seven-fold
  disclinations as indicated by red and blue dots. The zoomed-in plots of the
  highlighted isolated quadrupole and quadrupole pile are presented and analyzed
  in (a) and (b). $\epsilon_0=10\%$, $v_0=0.3$, $t=7,500\tau_0$, and
  $(p,q)=(39,0)$.  The lower panels show the histogram of $\theta$ (the angle
  between the $\vec{q}$ vector and the $x$ axis) at $\epsilon_0=10\%$ (as
  indicated by the red arrow in the right panel), and the plot of the relative
  number of horizontal quadrupoles ($\gamma$) versus the initial strain
  $\epsilon_0$. The statistical analysis is based on the sampling during $t\in
  [500\tau_0, 10,000 \tau_0]$ at the resolution of $\delta t=200\tau_0$.  (b)
  Demonstration of the formation of a quadrupole via the flip of the geometric
  bond from horizontal ($AB$ in configuration $H$) to vertical ($C'D'$ in
  configuration $V$). (c) Illustration of the connection of the quadrupole pile
  and the shear band. (d) Analysis of the lifespan $\tau$ of quadrupoles at
  $\epsilon_0=0$ (green), $\epsilon_0=5\%$ (blue), and $\epsilon_0=10\%$
  (black). The histograms of lifespans are presented in the lower panels.
  $v_0=0.3$. The sampling is during $t\in [40\tau_0, 100 \tau_0]$ with a
  temporal  resolution of $\delta t = 0.02\tau_0$, which is much shorter than
  the quadrupole lifespan. Note that the initial sampling time is chosen to be
  one or two orders of magnitude longer than the relaxation time to ensure that
  the velocity distribution is fully equilibrated.  
  }
\label{q}
\end{figure*}


\section{Results and discussion}

\subsection{Thermalization and emergence of quadrupoles}

We first create a thermal environment by disturbing the particle configuration
in mechanical equilibrium. Fig.~\ref{model}(b) shows a typical instantaneous
speed distribution that can be well fitted by the Maxwell distribution.
The thermalization process in the velocity distribution is
characterized by $\delta f(t)$, the deviation from the Maxwell distribution:
\begin{eqnarray}
  \delta f(t) = \frac{\int (f_t(v)-f_0(v))^2 dv}{\int f_0(v)^2 dv},
\end{eqnarray}
where $f_t(v)$ is the instantaneous speed distribution at time $t$, and $f_0(v)$
is the Maxwell distribution. Here, $\delta f(t)$ is used as the
operational equilibration criterion. Investigation of the lattices under the
disturbance of varying strength shows the uniform rapid relaxation of the speed
distributions within about two characteristic times, as shown in the inset
panel of Fig~\ref{model}(b).  From the perspective of microscopic dynamics, the
geometric nonlinearity arising from the triangular lattice that breaks
the integrability of the system is crucial for the realization of the thermal
equilibrium state; the thermalization process is characterized by the
proliferation of dynamical
modes~\cite{de2009relating,PhysRevLett.122.024102,PhysRevE.101.032211,yao2023non}.

The relation of the initial speed $v_0$ and the temperature (as derived from the
Maxwell distribution) at typical values of $\epsilon_0$ is shown in
Fig.~\ref{model}(c).  The deviation of the red curve ($\epsilon_0=-20\%$) is due
to the formation of folds; the released elastic energy is converted into the
thermal energy, raising the temperature. In the thermalized crystalline lattice,
the bond length is subject to fluctuation; see Fig.~\ref{model}(d). The
dispersion $\delta b_{\parallel}=\sqrt{k_B T}$, according to the energy
equipartition theorem. For the case in Fig.~\ref{model}(b), $\sqrt{k_B T}\approx
0.16$, which agrees with the simulation result.

To fully understand the impact of thermal agitation, we shall examine the
dynamics of individual particles. The Delaunay triangulation serves as a proper
tool to identify topological defects in instantaneous particle configurations,
which hold the information on relative positions of
particles~\cite{nelson2002defects}.  Specifically, the particles are
connected by geometric bonds according to the rule of the Delaunay
triangulation. Note that the established geometric bonds via Delaunay
triangulation can be distinct from the physical bonds (springs).  From the
established geometric bonding network, we can identify topological defects. In
the triangulated two‑dimensional lattice, the elementary topological defects
are $n$-fold disclinations, defined as particles surrounded by $n$ nearest
neighbors with $n \neq 6$~\cite{nelson2002defects}.


Figure~\ref{q}(a) shows the emergence of defects in a pre-stretched lattice
under the combined effects of stress and thermal fluctuation.  The red and blue
dots indicate five- and seven-fold disclinations identified by Delaunay
triangulation. A five-fold (or seven-fold) disclination in a triangular lattice
can be constructed by removing (or inserting) a wedge of angle $\pi/3$, which
leads to the convergence (or divergence) of originally parallel crystalline
lines passing through the
disclination~\cite{nelson2002defects,irvine2010pleats,yao2017topological}.
According to continuum elasticity theory, disclinations of opposite signs
attract and like signs repel in analogy to electric
charges~\cite{Nelson1987,nelson2002defects}. A pair of five- and
seven-fold disclinations constitute a dislocation, and four oppositely charged
disclinations form a quadrupole, in direct analogy to electric dipoles and
quadrupoles.

A pair of isolated dislocations appear on the upper right corner in
Fig.~\ref{q}(a).  Geometrically, in a triangular lattice an isolated dislocation
is created by inserting an array of particles; this is demonstrated by the green
and blue lines around the pair of dislocations
Fig.~\ref{q}(a)~\cite{Landau1986}.  In other words, the concept of dislocation
can be used to characterize the deformation whose effect is to insert an array
of particles into a triangular lattice. To further illustrate this point, we
construct a polygonal loop (in red) around the lower dislocation in the
dislocation pair in Fig.~\ref{q}(a); the side length of the polygon is two
lattice spacings.  The presence of the dislocation (or, equivalently, the
inserted particle array) prevents the closure of the loop. Quantitatively, by
treating the lattice as a continuum medium, the displacement vector $\vec{u}$
receives a certain finite increment $\vec{b}$ around any closed contour
enclosing a dislocation: 
\be
\oint du_i = -b_i, \label{dislocation}
\ee
where $\vec{b}$ is known as the Burgers vector~\cite{Landau1986}. $i=x,y$. According to
their geometric constructions, both disclinations and dislocations in a triangular
lattice cannot be removed by continuous deformation, thus classifying them as
topological defects.

We further analyze the quadrupole, which consists of four disclinations of
opposite signs organized in a square configuration. In Fig.\ref{q}(b), we show
the zoomed-in plot of the isolated quadrupole highlighted in green in
Fig.\ref{q}(a). The orientation of the quadrupole is specified by the
$\vec{q}$ vector, which connects the pair of five-fold disclinations. Unlike
disclinations or dislocations, quadrupoles can be removed by continuous
deformation. Due to the opposite Burgers vectors of the dislocations forming a
quadrupole, the contour integral enclosing a quadrupole in
Eq.(\ref{dislocation}) returns zero, as demonstrated by the closed polygonal
loop around the isolated quadrupole in Fig.\ref{q}(a). It implies that the
presence of a quadrupole does not disrupt the topological structure of the
lattice. To further understand that a quadrupole can be removed by continuous
deformation, we will now examine its formation at the microscopic level within a
triangular lattice.


In Fig.\ref{q}(b), we demonstrate the formation of a quadrupole via the flip of
the geometric bond from horizontal ($AB$ in configuration $H$) to vertical
($C'D'$ in configuration $V$)~\cite{supp_thermal}.  Consequently, $A'$ and $B'$
become five-fold disclinations; $C'$ and $D'$ become seven-fold disclinations
accordingly. The critical condition for the emergence of the
quadrupole (or the flip of the geometric bond) is determined by the rule of the
Delaunay triangulation~\cite{nelson2002defects}. In the Delaunay triangulation
algorithm, the construction of geometric bonds conforms to the principle of
maximizing the minimum angle in the triangulated configuration. In the
following, we analyze the deformation of configuration $H$ in Fig.\ref{q}(b)
consisting of isosceles triangles $\triangle ABC$ and $\triangle ABD$, and apply
the maximization principle of the minimum angle to derive for the critical
condition.

Under the horizontal stretching of configuration $H$ belonging to the elastic
lattice, the horizontal line $AB$ is elongated and the vertical line $CO$
simultaneously shrinks due to Poisson's effect. It is assumed that both
$\triangle ABC$ and $\triangle ABD$ remain isosceles triangles in the
deformation process. Consequently, the angle $\phi$ between $AB$ and $AC$ is
decreased with the horizontal stretching of configuration $H$. According to the
maximization principle of the minimum angle in the Delaunay triangulation, the
flip of the geometric bond $AB$ occurs when the minimum angle in configuration
$V$ becomes larger than that in configuration $H$. The minimum angle in
$\triangle ABC$ is $\phi$ for the case of triangular lattice subject to
horizontal stretching. The minimum angle in $\triangle A'C'D'$ is either $\angle
A'C'D'$ (which is $\pi/2-\phi$) or $\angle C'A'D'$ (which is $2\phi$). For the
former case, it is required that $\pi/2-\phi > \phi$ and $\pi/2-\phi < 2\phi$,
which lead to the inequality: $\pi/6 < \phi < \pi/4$. For the latter case,
$2\phi>\phi$ and $2\phi < \pi/2-\phi$, which lead to $\phi < \pi/6$.  To
conclude, the critical condition for the flip of the geometric bond from
configuration $H$ to $V$ is established as:
\be
\phi < \f{\pi}{4}.\label{cri}
\ee
According to Eq.(\ref{cri}), a quadrupole is formed when the value of $\phi$ is
reduced to be less than $\pi/4$ in thermalized, stretched lattice of zero tilt
angle. The resulting quadrupole structure serves as an important indicator for
the local fluctuation of the strain. The quadrupoles also represent the first
batch of thermally excited defects, and they constitute the precursors for the
subsequent isolated dislocations and disclinations via fission events at higher
temperatures~\cite{halperin1978theory,strandburg1988two}.


We further discuss the critical condition in terms of the amount of horizontal
stretching, above which the quadrupole is formed. In the stress-free crystalline
lattice of zero tilt angle, $\phi=\pi/3$ and $\overline{AB}=\ell_0$. Under the horizontal
stretching, the length of line $AB$ becomes $\ell_0(1+\epsilon)$, where
$\epsilon$ is the horizontal strain. Due to
Poisson's effect, the length of the vertical line $CO$ shrinks from
$\sqrt{3}\ell_0/2$ (in the stress-free state) to $(1-\sigma \epsilon) \times
\sqrt{3}\ell_0/2$ (in the deformed state).  According to the critical condition
in Eq.(\ref{cri}), $\tan\phi = \overline{CO}/\overline{AO}  <1$, which leads to
\be
\epsilon > \f{\sqrt{3}-1}{1+\sqrt{3}\sigma}. \label{cri_Gamma}
\ee
It is important to point out that the Poisson's ratio $\sigma$ is the only
elastic parameter entering the critical condition. This could be understood by
the physical meaning of $\sigma$. By its definition,
$\sigma=-\epsilon_{\perp}/\epsilon_{\parallel}$, where $\epsilon_{\parallel}$
and $\epsilon_{\perp}$ are the longitudinal and transverse strains,
respectively. As such, the quantity $\sigma$ captures the variation of the angle
$\phi$ arising from horizontal stretching, and is therefore linked to the
critical condition for bond flipping. Specifically, for a given horizontal
stretching, increasing $\sigma$ enhances the transverse compression and thereby
lowers the critical value of $\epsilon$ required to induce a quadrupole, which
is consistent with Eq.(\ref{cri_Gamma}).

For a harmonic triangular lattice fabricated by linear springs, $\sigma=1/3$,
which returns the critical value $\epsilon_c=2\sqrt{3}-3\approx 0.46$ according to
Eq.(\ref{cri_Gamma})~\cite{PhysRevA.38.1005,berinskii2017plane}. It means that
in an initially stress-free crystalline lattice of zero tilt angle, a quadrupole
emerges when a horizontal bond is stretched by an amount of $0.46\ell_0$, either
by mechanical deformation (at zero temperature) or under thermal agitation.

\begin{figure*}[t]  
\centering 
\includegraphics[width=7in]{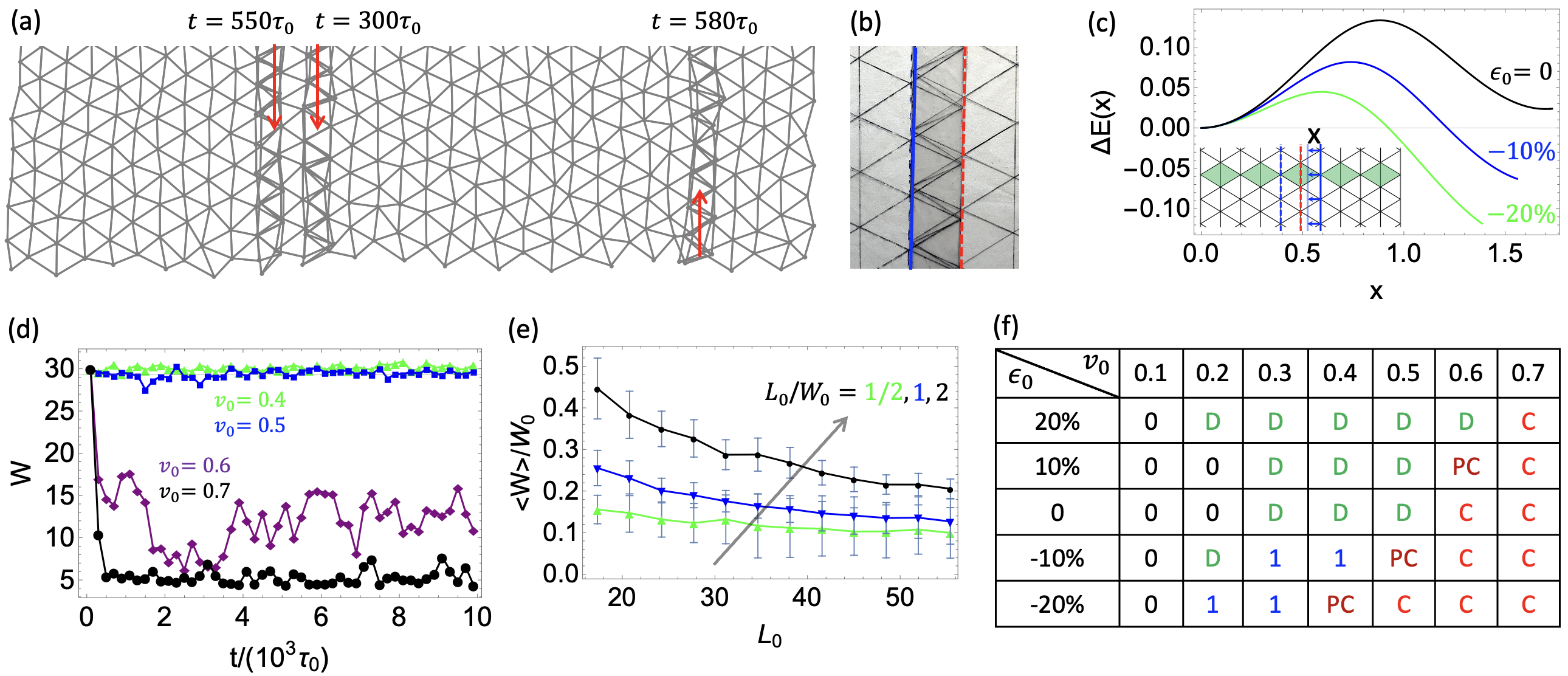}
  \caption{Thermally driven fold structure and the fold-driven collapse of the
  crystalline lattice. (a) Identification of the fold structure in thermalized,
  pre-compressed lattice. $\epsilon_0=-20\%$, $v_0=0.2$, $t=1,000\tau_0$, and
  $(p,q)=(20,20)$. The red arrows indicate the growth direction of the folds
  initially appearing on the edges at the indicated times. The upper part of the
  lattice is removed for visual convenience.  (b) The paper model to illustrate
  the formation of the fold via folding with respect to the reference line (in
  red). (c) The energy barrier during the associated quasi-static mechanical
  deformation process characterized by the displacement $x$, as demonstrated in
  the inset figure. $\Delta E$ is the variation of the elastic energy of the
  highlighted rhombs in the inset figure.  (d) Collapse of the lattice as
  characterized by its sudden shrinking under strong thermal agitation.
  $\epsilon_0=0$. (e) Plot of the relative time-averaged width $\la W \ra/W_0$
  versus $L_0$ at typical aspect ratio $L_0/W_0$. $\epsilon_0=0$, $v_0=0.7$, and $p=q$. (f) The phase diagram for the dynamical
  states of the system.  The symbol ``0" refers to the fold-free and defect-free
  state, "D" for the state with defects, ``1" for the folded state,
  "C" for the collapsed state [$\la W\ra/W_0<0.5$], and "PC" for the
  partially collapsed state [$\la W\ra /W_0 \in (0.5, 0.9)$]. $(p, q) = (20,
  20)$.  $W_0=30\ell$. In (e) and (f), the sampling is during $t\in
  [500\tau_0, 10,000 \tau_0]$ at the resolution of $\delta t = 100\tau_0$.
  }
\label{fold}
\end{figure*}

The critical condition in Eq.(\ref{cri_Gamma}) indicates that a finite amount of
energy is required to excite a quadrupole in an initially stress-free triangular
lattice. The excitation energy $\Delta E$ can be estimated based on the
deformation from configuration $H$ to $V$ in Fig.\ref{q}(b). By collecting the
elastic energies associated with the five springs in configuration $H$, we
obtain
\be
\Delta E = \f{ k_0 \ell_0^2  }{(1+\sqrt{3}\sigma)^2}(\gamma_0 - \gamma_1 \sigma +
\gamma_2 \sigma^2), \label{cri_e}
\ee
where $\gamma_0=7-\sqrt{3}-2\sqrt{6}\approx 0.37$,
$\gamma_1=6\sqrt{2}-6-4\sqrt{3}+2\sqrt{6} \approx 0.46$, and
$\gamma_2=9-6\sqrt{2}\approx 0.51$. $\Delta E$ represents the elastic energy
stored in a quadrupole. As such, quadrupoles act as stress absorbers in the
lattice subject to fluctuations. For an isotropic elastic medium composed of
linear springs in triangular lattice, $\sigma=1/3$~\cite{PhysRevA.38.1005}.  We
therefore have $\Delta E = k_0 \delta \ell^2$, where $\delta \ell \approx
0.33\ell_0$. The decreasing curve of $\Delta E$ as a function of $\sigma$ is
presented in SM~\cite{supp_thermal}.  Note that $\Delta E$ in Eq.(\ref{cri_e})
provides an underestimate of the quadrupole excitation energy, considering that
the configurational change from $H$ to $V$ also involves the deformation of
adjacent bonds. The concentration of elastic energy on the quadrupole is also
shown by the following static scaling analysis.  For a lattice of size $L$
($L\gg \ell_0$), the strain within the lattice caused by the local stretching
$\delta a$ at the quadrupole scales as: $\epsilon \sim \delta a/L \ll \delta
a/\ell_0$, the strain at the quadrupole.  Since the elastic energy density is
proportional to the square of local strain, the elastic energy is concentrated
on the quadrupole. Here, the quadrupole structure, identified by Delaunay
triangulation, concentrates or absorbs elastic energy, revealing its mechanical
impact.


An important observation is the stretch-driven alignment of thermally driven
quadrupoles in the lattice of zero tilt angle.  In the left panel below
Fig.\ref{q}(a), we show the distribution of $\theta$ (the angle between the
$\vec{q}$ vector and the $x$ axis) in a typical stretched lattice at
$\epsilon_0=10\%$. It is found that most quadrupoles are parallel to $x$ axis in
the stretched lattice of zero tilt angle.  The right panel further shows the
increase of the relative number of horizontal quadrupoles ($\gamma$) under
stronger initial stretching.  Even for the stress-free case ($\epsilon_0=0$),
the value of $\gamma$ is larger from $1/3$, indicating the preference of
alignment of quadrupoles along the $x$ axis.

According to the microscopic mechanism for quadrupole formation, the
observed horizontal alignment of quadrupoles in lattices with zero tilt angle
results from local horizontal bond stretching, which is thermally driven and
enhanced under pre-stretched condition. Note that even for the stress-free case
($\epsilon_0=0$), the cylindrical constraint with stress-free edges promotes
horizontal bond stretching over vertical bond stretching in thermal
fluctuations. In lattices of nonzero tilt angle, a common observation is that
the $\vec{q}$ vectors tend to align along the principal axis closest to the $x$
axis, due to the discrete nature of the quadrupole
orientation~\cite{supp_thermal}.

We also observe piling of quadrupoles along the principal axis of the lattice.
In Fig.\ref{q}(c), we show a typical pile of three quadrupoles highlighted in
Fig.\ref{q}(a). The formation of a quadrupole pile involves coordinated tilting
and stretching of the physical bonds (highlighted in cyan); in contrast, the
formation of an isolated quadrupole is caused by the stretching of a single
bond. The bonds in cyan are uniformly stretched~\cite{supp_thermal}.
Geometric analysis of the deformations of these bonds reveals the connection of
the quadrupole pile and shear deformation.  Specifically, the change of the
angle $\delta \phi$ between each pair of cyan and orange bonds in the
deformation is related to the shear strain: $u'_{xy} = (1/2)\tan\delta
\phi$~\cite{fung_elasticity}. $u'_{xy}$ is the strain tensor in the rotated
Cartesian coordinates $(x',y')$; the $x'$ axis is parallel to the quadrupole
pile~\cite{supp_thermal}. Based on the collected data $\delta \phi_i$, the
magnitude of the shear strain along the quadrupole pile is estimated as:
$|u'_{xy}|=0.15\pm
0.08$.  This thermally driven shear strain is much larger than the shear
strain $|u'^{(0)}_{xy}|$ created by the mechanical stretching along the
quadrupole pile at zero temperature; $|u'^{(0)}_{xy}| \approx
0.06$~\cite{supp_thermal}. Here, we shall emphasize that the quadrupole-pile
structure encodes the information of coordinated bond deformations under thermal
agitation, and it can thus be regarded as an embodiment of the transient
fluctuation of localized shear deformation.


Numerical approach allows us to track the transient dynamics of individual
quadrupoles. Simulations show that quadrupoles emerge and anchor at random sites
for a short duration (compared with the characteristic time $\tau_0$).
The presence of anchored quadrupoles in the range from $v_0=0.25$ to
$v_0=0.325$ offers a suitable window for measuring quadrupole lifespans.
The anchoring behavior can be attributed to the suppressed glide motion of the
pair of dislocations composing the quadrupole due to the balanced Peach–Koehler
forces; an anchored quadrupole in space represents the temporarily frozen
fluctuation of the strain~\cite{peach1950forces}. The dependence of the lifespan
$\tau$ of the quadrupole on $v_0$ is shown in Fig.\ref{q}(d); typical
distributions of lifespan $\tau$ are presented in the lower panels. On average,
an excited quadrupole lives for a duration of about $0.53\tau_0$ based on the
data in Fig.\ref{q}(d). The large fluctuations in the value of $\tau$ can be
understood in terms of the formation mechanism of a quadrupole; a quadrupole
vanishes upon a horizontal contraction to violate the critical condition.

Here, we resort to dimensional analysis to understand the mean lifespan of a
quadrupole. Since the emergence of quadrupoles results from the coordinated
thermal fluctuations and stress, the relevant physical quantities determining
quadrupole lifespan include temperature $T$, spring stiffness $k_0$ and the
dimensionless Poisson's ratio $\sigma$. The temperature $T$ is determined by the
quantities $m_0$ and $v_0$. We therefore use the combination of $m_0$, $v_0$ and
$k_0$ to construct a quantity with the dimension of time. It turns out that
$\tau = C m_0^{1/2}k_0^{-1/2}= C \tau_0$, where $C$ is a numerical factor of
order unity. The value of $C$ may vary with $\sigma$. $C=0.53$ by our numerical
experiment. Here, it is of interest to point out that dimensional analysis
suggests the independence of the lifespan on $v_0$. This is consistent with our
observation of the insensitivity of $\tau$ to the variation of $v_0$ over the
range investigated.  Physically, increasing $v_0$ raises the
temperature and leads to an isotropic stretching of the bonds. However, the
formation and persistence of a quadrupole require an anisotropic deformation
of the bonds; see Fig.~\ref{q}(b) for the formation mechanism of a quadrupole.
As such, while increasing $v_0$ generates more quadrupoles, it does not
necessarily affect the lifespan of any given quadrupole.

We also examine the lifespan of linear quadrupole piles. To observe
adequate emergence and annihilation events of linear quadrupole piles, we work
within the range of $v_0 \in [0.35, 0.425]$ at $\epsilon_0=0$, $10\%$, and
$20\%$. The observed linear quadrupole piles consist of two or three
quadrupoles. We find that the lifespan $\tau_p$ of a linear quadrupole pile
shows no significant dependence on $v_0$ or $\epsilon_0$ within the range
examined, consistent with the behavior of a single quadrupole.
$\tau_{p}/\tau_0=0.34 \pm 0.24$; a linear quadrupole pile thus has a shorter
mean lifespan than a single quadrupole ($\la\tau\ra/\tau_0=0.53$). Our statistical
analysis is based on the sampling during $t\in [40\tau_0, 50\tau_0]$ at the
resolution of $\delta t = 0.02\tau_0$; $v_0 \in \{0.35, 0.4, 0.425\}$, and
$\epsilon_0 \in \{0, 10\%, 20\% \}$.


\subsection{Folds arising in the compressed lattice}

We proceed to explore the dynamical behavior of thermalized, pre-compressed
lattices. Under thermal agitation, the fold structure emerges to release the
compressional stress; typical vertical folds in the lattice of $(p, q)=(20,20)$
are shown in Fig.~\ref{fold}(a). The formations of the three folds are initiated
on the edges of the lattice at different times, and they extend vertically at
the average speed of $0.31$. The folding processes are mostly irreversible.
We survey the parameter space of $\epsilon_0 \in [-20\%,0]$ and
$v_0\in [0.1,0.5]$, and identify only one unfolding event at $\epsilon_0=-2.5\%$
and $v_0=0.5$, where a tilted fold formed at $t=7,100\tau_0$ is unfolded at
$t=9,100\tau_0$.

Here, we use the paper model in Fig.~\ref{fold}(b) to illustrate the formation
of the fold structure. Folding with respect to the reference line (in red) leads
to the three layers of bonds within the region of fold, which is consistent with
the fold structures in Fig.~\ref{fold}(a). Now, we analyze the
energetics of fold formation from the perspective of a quasi-static mechanical
deformation process. As illustrated in the inset panel in Fig.~\ref{fold}(c),
the quasi-static folding process is characterized by the displacement $x$. At
any given $x$, the lattice is in mechanical equilibrium. Increasing $x$ leads to
a uniform horizontal stretching of the lattice, as well as the vertical
shrinking due to Poisson's effect. Let $\Delta E$ be the variation of the
elastic energy of the highlighted rhombs in the inset panel in
Fig.~\ref{fold}(c)~\cite{supp_thermal}. The plot of $\Delta E$ against the
displacement $x$ at typical values of $\epsilon_0$ is presented in
Fig.~\ref{fold}(c).

A common feature of the $\Delta E(x)$ curves in Fig.~\ref{fold}(c) is the
appearance of the energy barrier, indicating the suppression of fold formation
at zero temperature. Under a stronger pre-compression (less negative
$\epsilon_0$), a milder thermal fluctuation (smaller $v_0$) suffices to drive
the system across the energy barrier and trigger the fold formation.  As such,
the folds serve as stress releasers, facilitating the release of accumulated
compressional stress. The microscopic scenario of fold formation is as follows.
Once a short fold as a seed is initially formed on the edge, we observe the
coordinated motion of particles under the subtle interplay of thermal agitation
and compressional stress, leading to the smooth extension of the fold in the
lattice.


Under even stronger thermal agitation, the most important observation is the
collapse of both pre-stretched and pre-compressed lattices~\cite{supp_thermal}.
In Fig.~\ref{fold}(d), we show typical collapse dynamics in terms of the lattice
width at $v_0>0.5$. Microscopically, the collapse transition is realized by the
proliferation of horizontal or inclined folds; the vertical folds are suppressed
due to the cylindrical constraint. Here, the feature of self-intersection is
crucial for the collapse transition, which is similar to the crumpling
transition of a tethered phantom sheet in 3D
space~\cite{kantor1986statistical,Nelson2004c,JBOWICK2001255}.
Incorporating an energetic penalty for self‑intersection can delay the
onset of fold formation when thermal fluctuations are sufficiently strong. In
contrast, the quadrupole structure, which involves only mild bond deformation,
remains unaffected by the lattice's prohibition for self‑intersection.

We also explore the effect of lattice size on the collapsed state.
Figure~\ref{fold}(e) shows that it is even harder to collapse a thinner lattice
of given perimeter $L_0$ (larger $L_0/W_0$). At fixed aspect ratio, the degree
of collapse is enhanced with the increase of $L_0$. More information on the
characterization of the collapse of both pre-compressed and pre-stretched
lattices is presented in SM~\cite{supp_thermal}. Introducing initial strain
($\epsilon_0=\pm 20\%$) does not change the main conclusions obtained for the
stress-free cases shown in Fig.~\ref{fold}(e).


The phase diagram for the dynamical states of the thermalized, stressed
crystalline lattice is presented in Fig.~\ref{fold}(f). The degrees of thermal
agitation and stress are measured by the quantities $v_0$ and $\epsilon_0$. The
symbol ``0" refers to the fold-free and defect-free state, "D" for the state
with defects, ``1" for the folded state, "C" for the collapsed state, and "PC"
for the partially collapsed state. Figure~\ref{fold}(f) shows that the
characteristic quadrupole and fold structures define the distinct dynamical
states at varying levels of temperature and stress; the system is ultimately
collapsed under sufficiently strong thermal agitation. Comparison of
the last two rows of the phase diagram shows that the folded state emerges at a
smaller critical value of $v_0$ under stronger pre-compression condition
($\epsilon_0=-20\%$). This is consistent with the lower energy barrier in
Fig.~\ref{fold}(c) at $\epsilon_0=-20\%$, which is obtained from the
quasi-static analysis of the folding process.

To examine the robustness of the phase diagram, we further investigate
lattices of distinct tile angle and geometric size. Specifically, our analysis
focuses on the phase behavior of the $(p,q)=(39,0)$ lattice at zero tilt angle
and, for comparison, a reduced‑size lattice $(p,q)=(19,0)$ at varying $v_0$ and
$\epsilon_0$. The resulting phase diagrams for both cases are similar to that in
Fig.~\ref{fold}(f)~\cite{supp_thermal}. We also reduce the sampling interval
by half, and the resulting phase diagrams remain unchanged. In the phase
diagram, the collapsed states in the large-$v_0$ regime are subdivided into
partially (``PC") and fully (``C") collapsed states according to the specified
threshold $\la W\ra/W_0=0.5$. And the folded states (``1") and partially
collapsed states (``PC") are distinguished by the threshold $\la W\ra/W_0=0.9$.
Modest variations in these threshold values do not significantly alter the phase
diagram.

We finally discuss the triangular lattice model under harmonic interaction and
possible extensions of the current work.  The harmonic triangular lattice system
provides a general model for the dynamical behaviors of diverse 2D regular
particle packings near mechanical equilibrium, where the physical interaction
can be approximated by a harmonic potential. With the intrinsic geometric
nonlinearity and the resulting broken integrability, the harmonic triangular
lattice model serves as a particularly effective platform to explore the
thermalization problems from the dynamical perspective as demonstrated in this
work. While this work focuses on the Hamiltonian dynamics of an isolated
triangular lattice system, it is a natural extension to incorporate interactions
with the environment into the model. For example, it is worthwhile to
investigate the lattice system's dynamical response to mechanical deformations
of the cylindrical substrate, which has a strong connection to the manipulation
of the stress regulators.  It is also of interest to model the impact of
environment as a source of dissipation and to analyze the resulting dissipative
dynamical behaviors of the stress regulators.

\section{Conclusion}

In summary, we have shown that a thermalized, stressed lattice system admits
the characteristic quadrupole and fold structures that arise from the intricate
interplay of thermal agitation and mechanical stress. As a tangible embodiment of
thermal fluctuations, these stress regulator structures offer a unique
perspective on the characterization and manipulation of the elusive
fluctuations. While the adaptation of two-dimensional crystal lattices to curved
geometries and mechanical constraints has been extensively explored from both
elastic and plastic perspectives in soft matter
physics~\cite{cerda2002wrinkling,bowick2009two,holmes2010draping,audoly2010elasticity,vernizzi2011platonic,Grason2013,wales2014chemistry,Mitchell2016,sun2025defect},
the inquiry into the thermalized stressed lattice system in this work suggests
the avenue of re-examining classical mechanical systems subject to thermal
agitation.  Especially, the thermal physics of stress regulators, whose roles in
mechanical systems have long been studied, remains a promising area awaiting
further explorations~\cite{Landau1986,audoly2010elasticity}.
\\

\section{Acknowledgements}

This work was supported by the National Natural Science Foundation of China
(Grants No. BC4190050).


%

\end{document}